\documentclass[aps,prc,twocolumn,groupedaddress,showpacs]{revtex4}

\usepackage{graphicx}

\begin{document}

\title{Reply to ``Comment on `Symmetry energy and the isospin dependent equation of state' "}
\author{D.V. Shetty, S.J. Yennello, A.S. Botvina\footnote{On leave from Institute for Nuclear Research,
117312 Moscow, Russia.}, G.A. Souliotis, M. Jandel, E. Bell, A. Keksis, S. Soisson, B. Stein, and J. Iglio}
\affiliation{Cyclotron Institute, Texas A$\&$M University, College Station, Texas 77843, USA}
\date{\today}

\begin{abstract}
We show that the large sequential decay corrections obtained by Ono {\it et al} [nucl-ex/0507018], is in 
contradiction with both the other dynamical and statistical model calculations carried out for the same systems and energy.  
On the other hand, the conclusion of Shetty {\it {et al.}} $[$Phys. Rev. C 70, 011601R (2004)$]$,  that the 
experimental data favors Gogny$-$AS interaction (obtained assuming a significantly smaller sequential decay effects), 
is consistent with several other independent studies.  
\end{abstract}


\maketitle

In Ref. \cite{SHE04}, we compared the experimentally determined isoscaling parameter $\alpha$, to those 
of the dynamical AMD calculation of Ono {\it {et al.}} \cite{ONO03}. The comparison was carried out without 
the sequential decay corrections. It was shown that the experimental data agrees better with the choice of Gogny-AS 
interaction. The conclusion was based on the assumption that the sequential decay effect is small and less than 15$\%$ 
for the systems studied (in fact, a correction of 15$\%$ results in even better agreement with Gogny-AS interaction). The 
same data was also compared with a statistical model calculation in Ref. \cite{SHE05}, and the sequential decay effect 
was observed to be small. The magnitude of this effect was found to be similar to those assumed in the dynamical model 
comparison in Ref. \cite{SHE04}. 
\par
\begin{figure}
\includegraphics[width=0.50\textwidth,height=0.45\textheight]{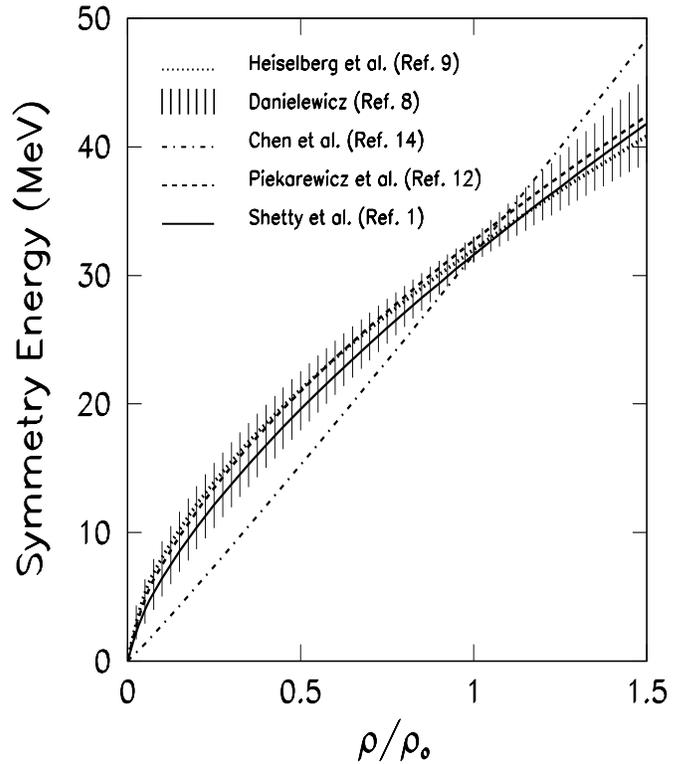}
\caption{Comparison of the density dependence of the symmetry energy obtained from various different studies.
See text for details.}
\end{figure} 

\begin{table*}
\caption{\label{tab:table1}Parametrized form of the density dependence of the symmetry energy obtained
from various independent studies.}
\begin{ruledtabular}
\begin{tabular}{cccccccc}
   Reference                             &  &   Parametrization                          &  Studies                             \\
\hline
Heiselberg {\it {et al.}} \cite{HEI00}   &  &  32.0($\rho$/$\rho_{o}$)$^{0.60}$          &  Variational calculation             \\
Danielewicz {\it {et al.}} \cite{DAN04}  &  &  31(33)($\rho$/$\rho_{o}$)$^{0.55(0.79)}$  &  BE, skin, isospin analog states     \\
Chen {\it {et al.}} \cite{CHE05}         &  &  31.6($\rho$/$\rho_{o}$)$^{1.05}$          &  Isospin difussion                   \\  
Piekarewicz {\it {et al.}} \cite{PIE05}  &  &  32.7($\rho$/$\rho_{o}$)$^{0.64}$          &  Giant resonances                    \\ 
Shetty {\it {et al.}}  \cite{SHE04}      &  &  31.6($\rho$/$\rho_{o}$)$^{0.69}$          &  Isotopic distribution            
\end{tabular}
\end{ruledtabular}
\end{table*}

Ono {\it {et al.,}} in their comment \cite{ONO06} show that the effect of sequential decay in the dynamical
AMD model is significantly large (about 50 $\%$), and therefore the conclusion of Shetty {\it {et al.,}} \cite{SHE04}, 
that fragment data show preference for the Gogny-AS interaction, cannot be justified. 
The conclusions of this comment are based on the premise that the secondary de-excitation effect for 
the dynamical model is large and significantly different from the statistical model calculations.
Recently, another dynamical calculation using IQMD model, has shown that no significant
difference exists between the primary and the secondary $\alpha$ \cite{TIA06}, thereby indicating that the
effect of sequential decay is very small. The observed small difference between the primary and the secondary
$\alpha$ is in direct contradiction with the large sequential decay corrections reported by Ono {\it {et al.,}} \cite{ONO06}. The
sequential decay effect from the IQMD calculation was carried out for the same systems and beam energy as studied 
by Ono {\it {et al.}} \cite{ONO06} using the AMD model. 
\par
The stark contrast in the sequential decay effect between two different dynamical calculations, as well the
statistical calculation, for the same systems and energy, is not clear at this moment.
If one assumes that the sequential decay is determined by the excitation energy, charge (Z) and mass (A) of the fragments, 
and not by the process that leads to these fragments, the calculations of Ono {\it {et al.}} \cite{ONO06}
are also in conflict with their own sequential decay calculation from statistical models \cite{TSA01,TAN01}. 
It appears that the isoscaling parameter, like in all statistical calculations,
remains a robust observable against sequential decay even in dynamical calculations.
While this needs to be verified further, we show in the following section that   
by assuming a negligibly small sequential decay (as established in various statistical
calculations and recently shown in dynamical IQMD calculation), the results of Shetty
{\it {et al.}} \cite{SHE04}, that the data agrees better with the choice of Gogny-AS interaction, are consistent with several 
other independent studies. 
\par
Figure 1 shows the density dependence of the symmetry energy obtained from the work of Shetty {\it {et al.}} \cite{SHE04} 
along with those obtained from various other studies. The
shaded region in the figure correspond to the result of Danielewicz \cite{DAN04} obtained by constraining the binding energy, 
neutron skin thickness and isospin analog state in finite nuclei. The dotted curve correspond to the parametrization
adopted by Heiselberg {\it {et al.}} \cite{HEI00} in their studies on neutron stars, and obtained by fitting the
predictions of the variational calculations of Akmal {\it {et al.}} \cite{AKM98}. The dashed curve correspond to those
obtained recently from an accurately calibrated relativistic mean field interaction, for
describing the Giant Monopole Resonance (GMR) in $^{90}$Zr and $^{208}$Pb, and the IVGDR in $^{208}$Pb, by
Piekarewicz {\it {et al.}} \cite{PIE05,TOD05}. The solid curve correspond to those obtained from Gogny-AS
interaction that explain the results of Shetty {\it {et al.}} \cite{SHE04}, assuming a negligibly small
sequential decay effect. The dot-dashed curve correspond to
those used for explaining the isospin difussion data of NSCL-MSU \cite{TSA04} by Chen {\it {et al.}} \cite{CHE05}.  This 
dependence has now been modified to include isospin dependence of the in-medium nucleon-nucleon cross-section, and is in 
close agreement with that of Shetty {\it {et al.}} \cite{SHE04}. The parameterized form of the density dependence of
the symmetry energy obtained from all these studies are as shown in table I.  
\par
The close agreement between various studies shown in figure 1, suggests that the assumptions made and the 
conclusion arrived at in Shetty {\it {et al.}} \cite{SHE04}, are probably correct. 
\par
Finally, Ono {\it {et al.}} \cite{ONO06}, argues that the method of `` extrapolating " the asymmetry of the primary
fragments as done in Shetty {\it {et al.}} \cite{SHE04}, is problematic. They base their argument on estimation of $f$
$\sim$ 0.75 for $^{64}$Zn + $^{197}$Au, $^{64}$Zn + $^{92}$Mo and $^{64}$Zn + $^{58}$Ni reactions. 
 These systems are highly asymmetric and have charge and mass nearly twice that of the
systems studied by Ono {\it {et al}} \cite{ONO03}, and Shetty {\it {et al.}} \cite{SHE04,SHET05}.
The systems studied in Shetty {\it {et al.}}
\cite{SHE04, SHET05} ($^{58}$Fe + $^{58}$Fe, $^{58}$Fe + $^{58}$Ni, $^{58}$Ni + $^{58}$Ni, $^{40}$Ar + $^{58}$Fe, $^{40}$Ar +
$^{58}$Ni and $^{40}$Ca + $^{58}$Ni), and Ono {\it {et al.}} \cite{ONO03} ($^{40}$Ca + $^{40}$Ca, $^{48}$Ca + $^{48}$Ca and $^{60}$Ca +
$^{60}$Ca), are nearly similar, and it is therefore natural to interpolate as done in fig. 3
of Shetty {\it {et al.}} \cite{SHE04}. Ono {\it {et al.}}'s assertion that the calculations are strictly for Ca + Ca reactions
is contrary to the spirit of any theoretical model calculations.
\par
This work was supported in parts by the Robert A. Welch Foundation through grant No. A-1266, and the Department of Energy 
through grant No. DE-FG03-93ER40773.

\end{document}